\documentstyle[amssymb,aps,multicol,prl,epsf]{revtex}

\newcommand \beq  {\begin{equation}}
\newcommand \eeq  {\end{equation}}
\newcommand \bea {\begin{eqnarray} }
\newcommand \eea {\end{eqnarray}}

\begin{document}
\title{Application of the scattering rate sum-rule to the interplane optical conductivity of high
temperature superconductors: pseudogap and bi-layer effects}
\author{N. Shah and A. J. Millis}
\address{Center for Materials Theory\\
Department of Physics \& Astronomy, Rutgers University\\
136 Frelinghuysen Road, Piscataway, NJ 08854}
\maketitle

\begin{abstract}
We use a recently proposed model of the interplane conductivity of high
temperature superconductors to investigate the ``scattering rate sum-rule'' introduced by Basov and co-workers. We present a new derivation of the sum-rule. The quantal and thermal fluctuations of the order parameter which have been argued to produce the observed {\em pseudogap} behavior are shown to increase the total integrated ``scattering rate'' but may either increase or decrease the ``quasiparticle'' contribution from frequencies greater than twice the superconducting gap.
\end{abstract}


\begin{multicols}{2}


One of the crucial current questions in the field of high temperature
superconductivity is the physics of the {\em pseudogap} regime observed in
underdoped systems \cite{Millis00}. This regime is characterized by the
opening of a gap in the electronic excitation spectrum. \ The origin of the
gap is the subject of debate, with some authors attributing it to a density
wave of one sort or another \cite{density} and others attributing it to the
superconducting gap, with long ranged superconducting order prevented by
loss of phase stiffness \cite{Millis00,Emery95}. A second, possibly related, issue is the
degree to which the superconducting ground state differs from the familiar
{\em BCS} mean field state. In particular, the possibility of strong
superconducting phase fluctuations signaling proximity to a nontrivial
insulating phase has attracted considerable interest \cite{groundstate}.

Measurements of the optical conductivity are an
important source of information about the {\em pseudogap} regime. For example,
the absorptive part of the interplane ({\em c-axis}) conductivity shows a dramatic suppression at low frequencies as the {\em pseudogap} is formed. The detailed frequency and temperature dependence was recently
argued to support a superconducting origin of the {\em pseudogap} \cite
{Ioffe99,Ioffe00}. Very recently, Basov and co-workers have introduced a new
way of interpreting optical conductivities \cite{Basov00}, based on a 
``scattering rate sum rule'' involving the real part of the inverse of the
complex conductivity which is defined by   
\bea
 \tau^{-1} (\omega) = \mathop{\rm  Re} \sigma^{-1}(\omega) = \frac{\mathop{\rm  Re} \sigma}{\mathop{\rm Re}\sigma^2
+\mathop{\rm Im}\sigma^2} . 
\label{exp}
\eea
In this paper we provide a derivation of the sum rule which is convenient for our analysis and analyze the theory of refs\cite
{Ioffe99,Ioffe00} for {\em single-layer} high-T%
$_{c}$ superconductors such as $La_{2-x}Sr_{x}CuO_{4}$ and it's recent generalization \cite{SMI} to {\em bi-layer} materials such as 
$YBa_{2}Cu_{3}O_{6+x}$ in terms of the new scattering rate sum rule. We show that
the so-defined  ``scattering rate'' behaves in a nonintuitive manner, increasing
in the {\em pseudogap} regime. This behavior may be a valuable probe of the quantal and thermal
phase fluctuations of the superconducting order parameter.



We first present an alternative derivation of the scattering
rate sum-rule. We suppose that at large frequency and for any temperature, Re$\sigma ^{-1}(\omega \rightarrow \infty, T)\rightarrow \sigma _{0}^{-1}$ with $\sigma _{0}$ a constant independent of $\omega$ and T. In the familiar Drude model, $\sigma _{0}^{-1}$ is proportional to 
the scattering rate, while in realistic models electrons decouple from the underlying crystal at sufficiently high frequencies, so $\sigma
_{0}^{-1}=0$. This assumption implies that the inverse conductivity $\sigma^{-1}(\omega ,T)$ which is a causal response function, obeys a ``subtracted''
Kramers-Kronig relation, 
\begin{equation}
\mathop{\rm Im}\sigma ^{-1}(\omega )=\int_{-\infty }^{\infty }\frac{dx}{\pi }\frac{%
\mathop{\rm Re}\sigma ^{-1}(x)-\sigma _{0}^{-1}}{\omega -x}.  \label{KK}
\end{equation}
We use the $\omega \rightarrow \infty $ limit of Eq \ref{KK}  
to obtain an equation for the changes in the inverse
conductivity as the temperature is changed from $T_{1}$ to $T_{2}$,
\bea
&&\lim_{\omega \rightarrow \infty }\omega \left( 
\mathop{\rm Im}%
\sigma ^{-1}(\omega ,T_{1})-%
\mathop{\rm Im}%
\sigma ^{-1}(\omega ,T_{2})\right) \nonumber \\
=&&\int_{0}^{\infty }\frac{2dx}{\pi }%
\mathop{\rm Re}%
\left( \sigma ^{-1}(x,T_{1})-\sigma ^{-1}(x,T_{2}\right).  \label{tsumrule}
\eea
In many circumstances, including the c-axis conductivity of high temperature
superconductors, the integral on the right hand side converges rapidly with frequency.

We now apply this sum rule to a theoretical model recently proposed \cite
{Ioffe99,Ioffe00} for the {\em pseudogap} behavior of the c-axis conductivity of
underdoped high temperature superconductors. In these papers it is argued that the weakness of the interplane coupling allows one to treat the problem by perturbation theory in the interplane coupling, leading to an expression for the c-axis conductivity involving a convolution of normal and anomalous in-plane Green functions. The model has three
parameters: an in-plane scattering rate $\Gamma $, a superconducting gap $\Delta $, and a {\em Debye Waller} factor $\alpha$ characterizing the strength of in-plane thermal and quantal fluctuations of the phase of the superconducting order parameter. The observed $d$-wave structure of the
superconducting gap \cite{Kirtley} and the variation around the Fermi surface of  $\Gamma $ \cite{Valla99} may be neglected because the interplane coupling is only
appreciable in momentum regions in which both $\Gamma $ and $\Delta$ are large \cite{Andersen94,Chakravarty94}. Photoemission \cite{Valla99} and c-axis optical data show that in the relevant region of the Fermi surface,  $\Gamma $ is very large and has negligible temperature dependence. We take the $\Gamma \rightarrow \infty$ limit in what follows so that in the model of refs\cite{Ioffe99,Ioffe00} temperature enters via the values of $\Delta$ and $\alpha$. At high temperatures, above the {\em pseudogap} regime, $\Delta = \alpha = 0$. The {\em pseudogap} regime is characterized by a $\Delta \ne 0$ but $\alpha = 0$ (superconductivity destroyed by phase fluctuations). In the superconducting state $\alpha \ne 0$, and an important question is whether as $T \rightarrow 0$, $\alpha \rightarrow 1$ (as in {\em BCS} mean field theory) or $\alpha < 1$, signaling quantum fluctuations. The results of refs\cite{Ioffe99,Ioffe00} lead to  
\begin{equation}
\sigma (\omega \rightarrow \infty ,T)\rightarrow \sigma _{0}\left[
1+i\Delta \pi \left( \alpha -1\right)/ 2\omega +{\cal O}(1/\omega ^{2})\right]. 
\label{highfreq}
\end{equation}
Combining Eqs \ref{highfreq} and \ref{tsumrule} yields our final expression 
\bea
f(\alpha)& =&\int_{0}^{\infty }\frac{2dx}{\pi }%
\mathop{\rm Re}%
\left( \sigma ^{-1}(x,\Delta ,\alpha )-\sigma ^{-1}(x, \Delta =0,\alpha
)\right) \nonumber \\
&=&\frac{\Delta \pi \left( 1-\alpha \right) }{2\sigma _{0}}.
\label{finalsum}
\eea
Thus the opening of the {\em pseudogap} $(\Delta > 0, \alpha = 0)$ corresponds to an {\em increase} in the area under the Re$\sigma^{-1}$ curve, while the subsequent entry into the superconducting state $(\alpha > 0)$ leads to a {\em decrease}. 

The sum rule on $%
\mathop{\rm Re}%
\sigma ^{-1}$ is sometimes interpreted as a ``scattering rate sum rule'',
because it is conventional to write $\sigma $ in the `modified Drude form' $%
\sigma (\omega )=\omega _{p}^{2}/(-i\omega \lambda (\omega )+\gamma (\omega
))$ in which the real part of $\sigma ^{-1}$ is interpreted as a scattering
rate. Because the opening of the {\em pseudogap}, which corresponds to a decrease in the density of states, leads to an increase in the integrated area of the Re$\sigma^{-1}$ curve, care is required in using this interpretation of scattering rate sum rule data.


We now apply the results of \cite{Ioffe99,Ioffe00} to study in
more detail the behavior of $%
\mathop{\rm Re}%
\sigma ^{-1}$ in the {\em pseudogap} regime of a {\em single-layer} superconductor.  Fig 1 plots the contributions to  $\tau^{-1}=%
\mathop{\rm Re}%
\sigma ^{-1}(\omega ,\Delta ,\alpha )$ for the normal state and for superconducting state at $\omega >2\Delta$ for several values of $\alpha$. For $\alpha < 1$, $\Delta \ne 0$, there is an additional  delta function contribution to $\tau^{-1}$ arising from a zero-crossing of Im$\sigma$ at $\omega < 2\Delta$ (not shown in Fig 1). The dashed flat curve (labeled by n) corresponds to the normal
$(\Delta = 0)$ state, while the other curves
correspond to the superconducting state for different values of $\alpha $. One sees that the form and magnitude of $\tau^{-1}$ depend sensitively on the strength of the fluctuation parameter. For some $\alpha$-values (0.7 for example) the area under $\tau^{-1}$ for $\omega > 2 \Delta$ is manifestly greater than for the normal state, whereas for small $\alpha$ ($\alpha = 0.1$ for example) it is manifestly less. 

\vspace{0.25cm} 
\centerline{\epsfxsize=3truein \epsfbox{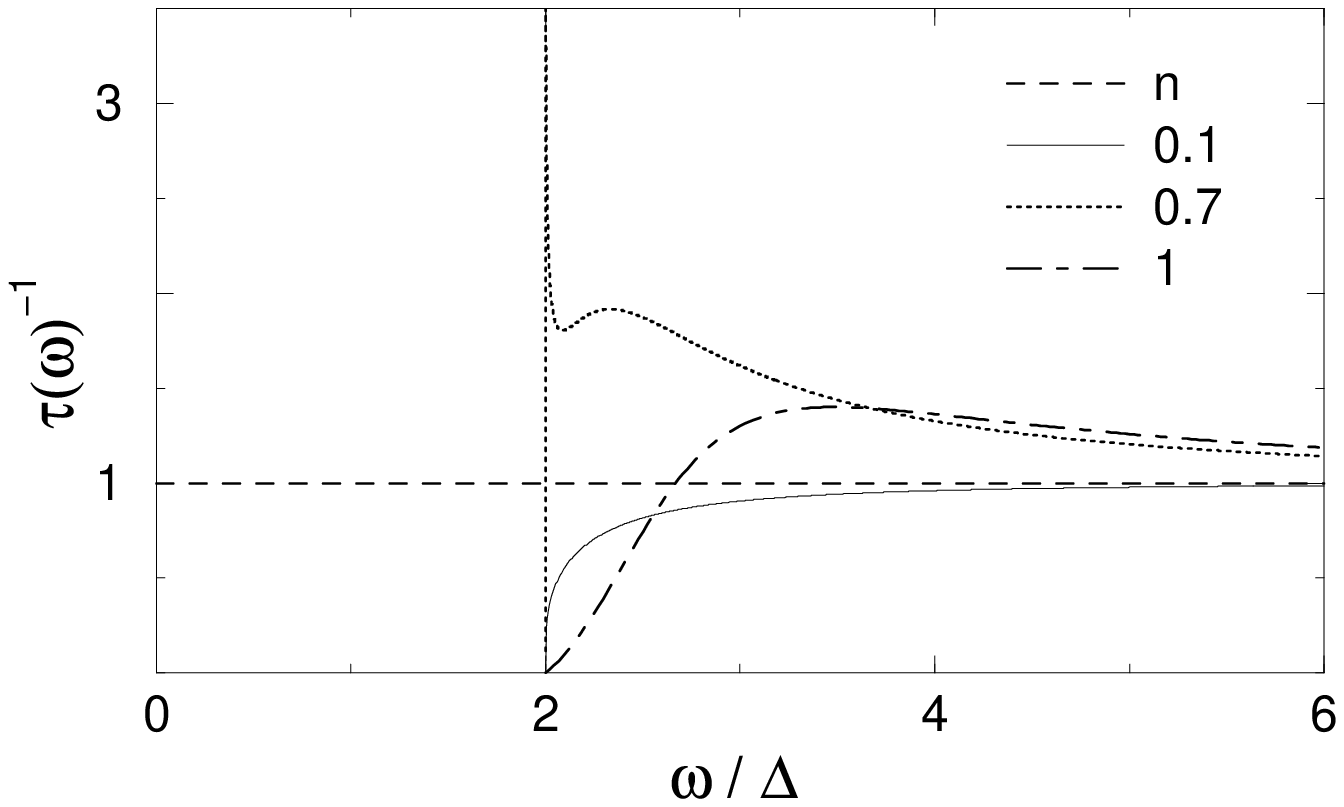}}
{\footnotesize \textbf{Fig. 1} Optically defined scattering rate $\tau^{-1}$ (Eq \ref{exp}) for {\em single-layer} superconductor normalized to $\omega \rightarrow \infty$ limit for normal state (dashed line) and for superconducting state with  different values of fluctuation parameters $\alpha = 0.1, 0.7$ and 1. For superconducting state only the contribution for $\omega >2\Delta $ is shown. }
\vspace{0.25cm}

To further characterize the structure in  $\tau^{-1}$ it is useful to define 
\bea
&&f_{\omega > 2 \Delta}(\alpha)=\nonumber\\ 
&& \int_{2 \Delta}^{\infty} 
\frac{2dx }{\pi }\mathop{\rm Re} \sigma^{-1}(x,\Delta ) - \int_{0}^{\infty} \frac{2dx }{%
\pi }\mathop{\rm Re} \sigma^{-1}(x,\Delta = 0 ) \label{sum-rule1} 
\eea
which excludes from Eq \ref{finalsum} the contribution coming from any $\omega < 2 \Delta$ delta functions. This quantity is shown in Fig 2 along with the normalized total area difference $f(\alpha)$. One sees that for $\alpha$ near 1 the increase in $\tau^{-1}$ comes mainly from the $\omega > 2 \Delta$ region, whereas as $\alpha \rightarrow 0$ there is a decrease in the  $\omega > 2 \Delta$ contribution to $\tau^{-1}$, but the increased value of the `pole' contribution at $\omega < 2 \Delta$ leads to a net increase in the sum. 

\vspace{0.25cm} 
\centerline{\epsfxsize=3truein \epsfbox{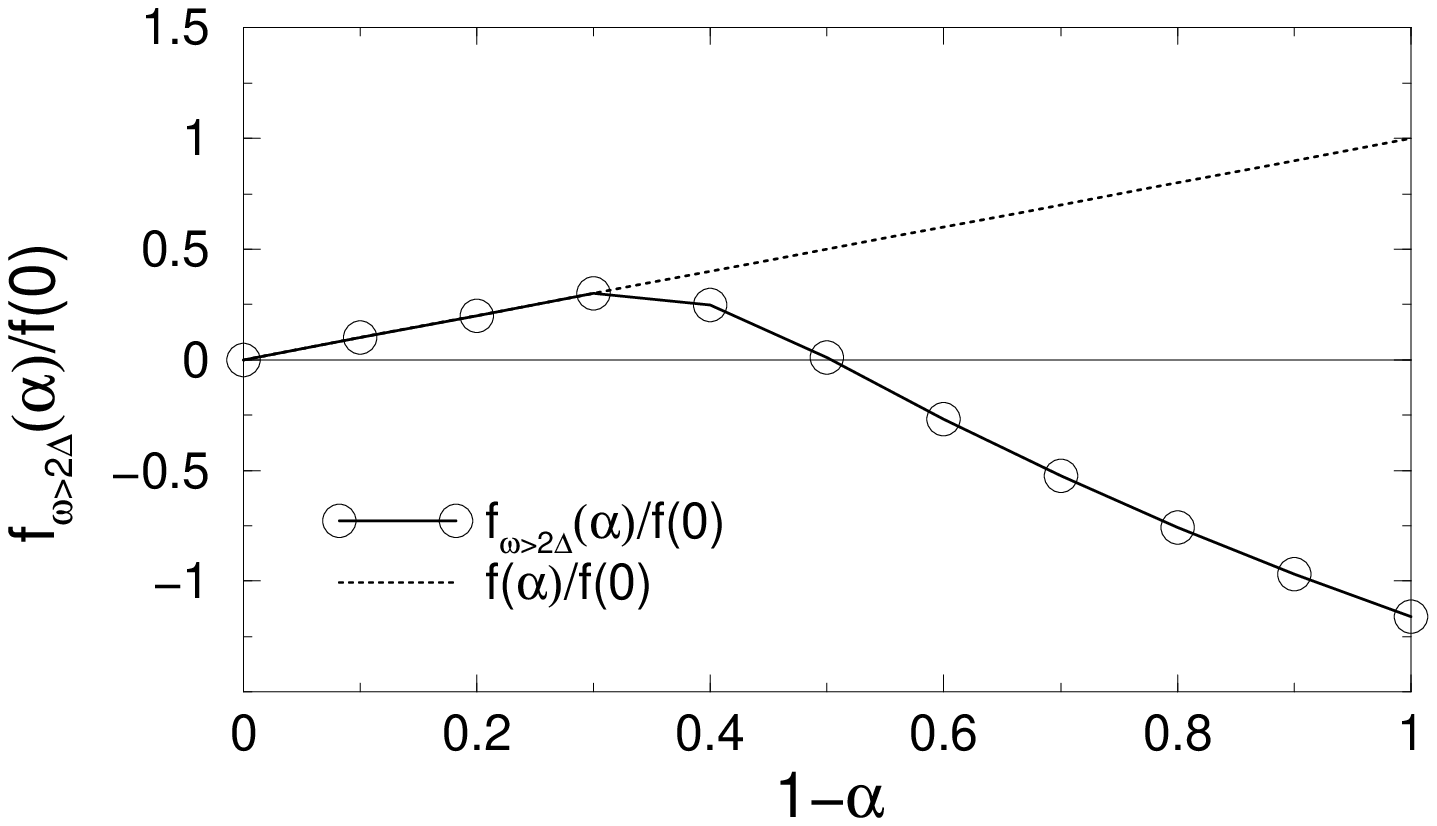}}
{\footnotesize \textbf{Fig. 2} Difference in the integrated area of $\tau^{-1}$ (Eq \ref{exp}) between  the gapped (superconducting or {\em pseudogap}) state and ungapped ($\Delta = 0$) state as function of fluctuation parameter $\alpha$. Open circles: quasiparticle ($\omega > 2 \Delta$) contribution only (Eq \ref{sum-rule1}). Dotted line : total contribution. Difference $f(\alpha) - f_{\omega > 2 \Delta}(\alpha)$ is greater than zero for all $\alpha < 1$, but is too small to be seen for $\alpha \ge 0.7$. }
\vspace{0.25cm}

To elucidate this behavior we show in Fig 3, the real and imaginary parts of the conductivity for $\alpha = 1$ (fully phase
coherent or {\em BCS}), $\alpha = 0.6$ and $0.1$ superconductor. For $\alpha = 1$ (Fig 3a) the  imaginary
part diverges as $1/\omega$ as $\omega \rightarrow 0$ and with increasing $\omega$  monotonically goes to zero (as $\omega ^{-3}$ at high frequencies). The real part increases
gradually from zero at the gap edge due to the type-II coherence factors. The $\tau^{-1}$ in this case is zero below the gap and is a
smooth function above the gap as shown in Fig 1. The weak maximum in $\tau^{-1}$ is close to the position of a minimum in Re$\sigma^2
+$ Im$\sigma^2$. 

\vspace{0.25cm} 
\centerline{\epsfxsize=3truein \epsfbox{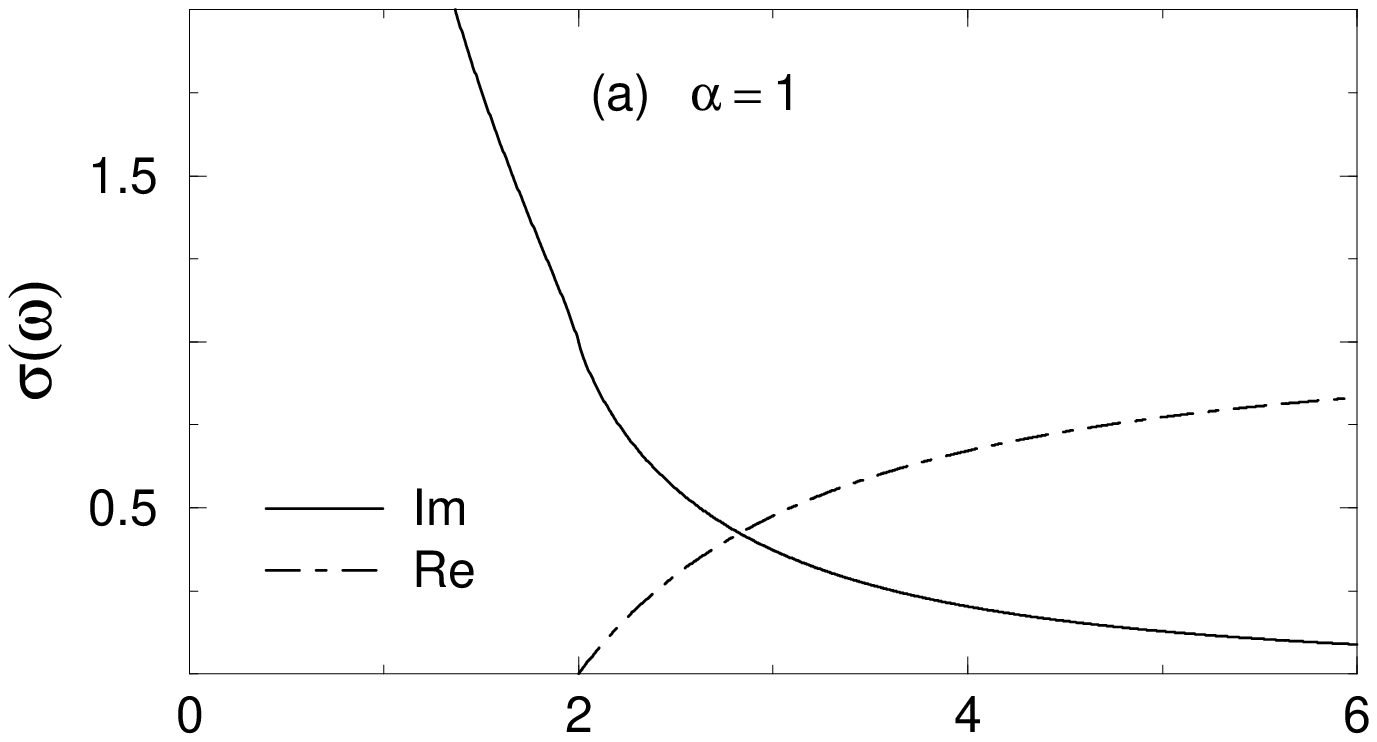}}
\vspace{0.01cm}
\centerline{\epsfxsize=3truein \epsfbox{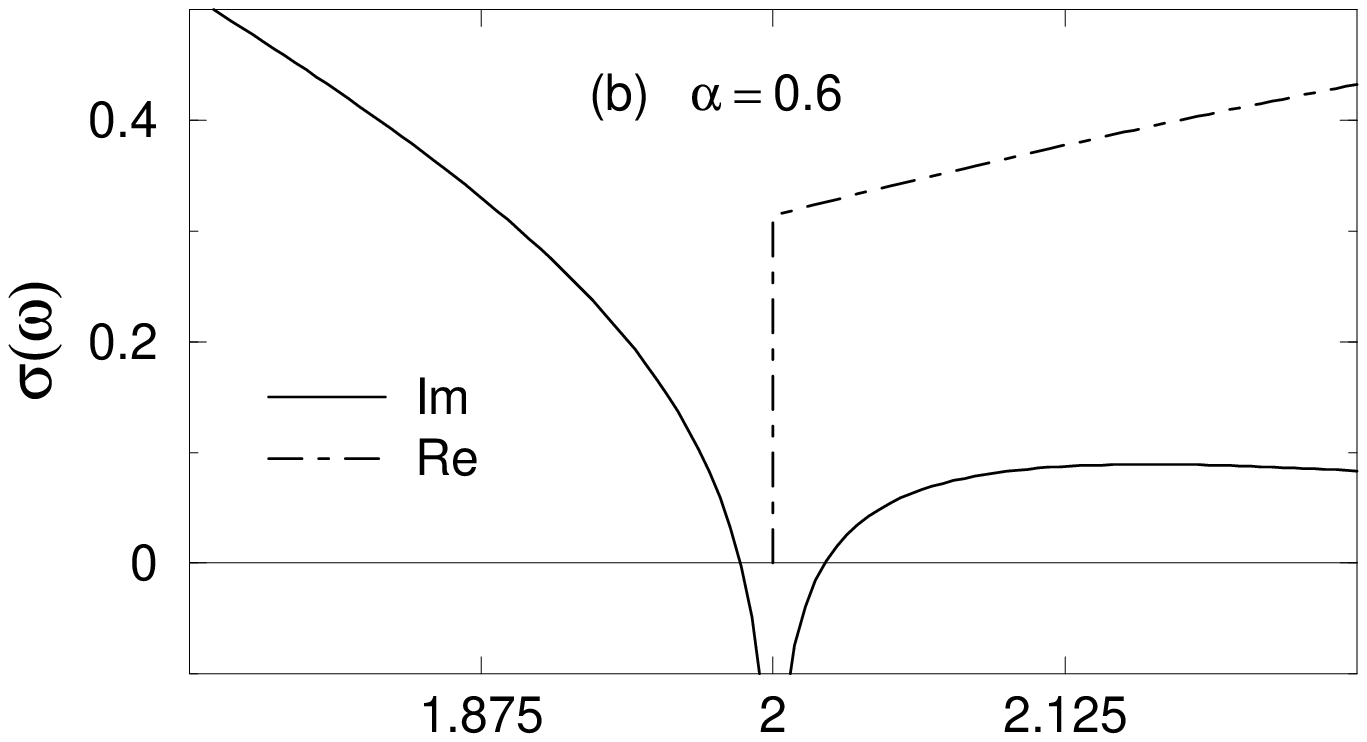}}
\vspace{0.01cm} 
\centerline{\epsfxsize=3truein \epsfbox{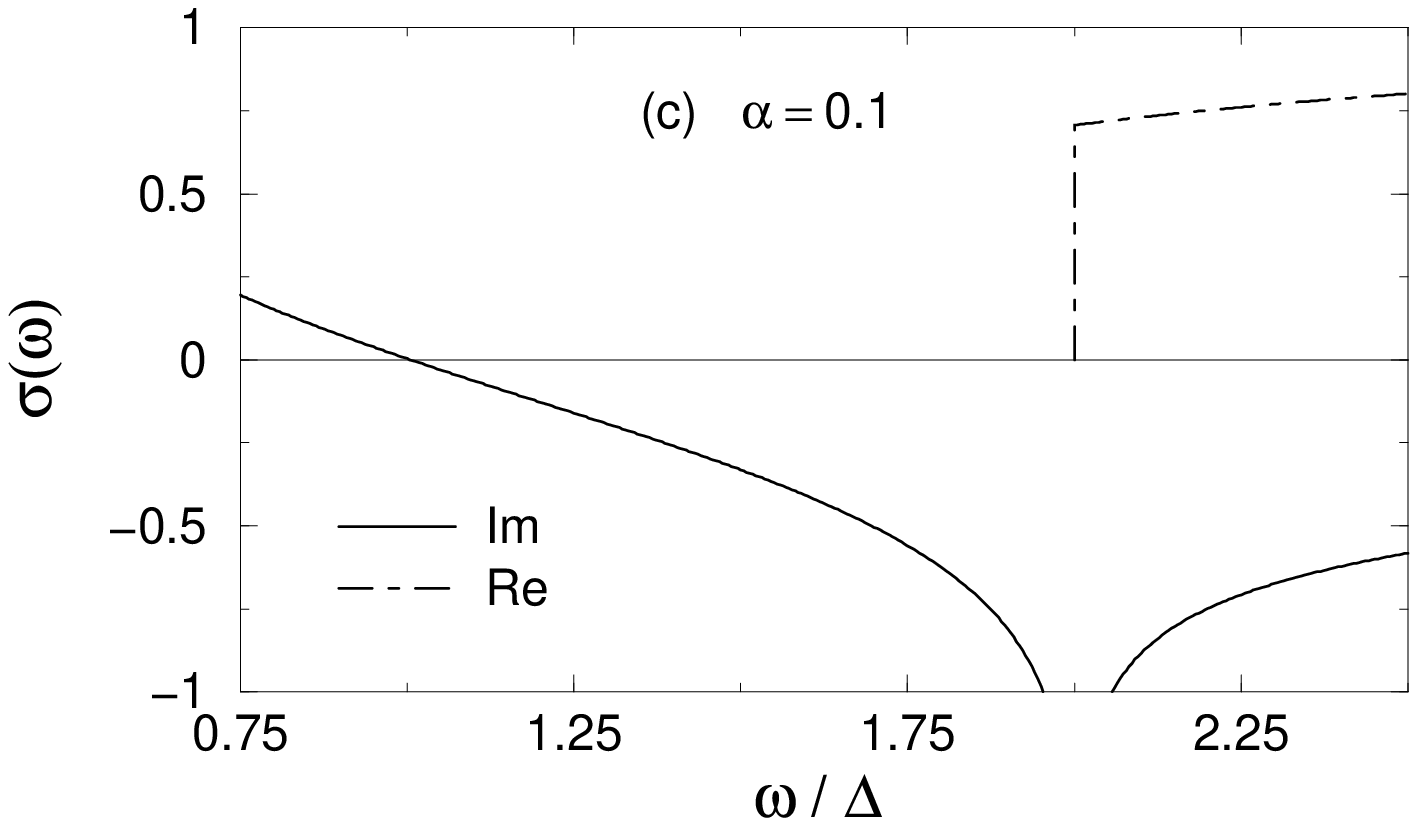}}
{\footnotesize \textbf{Fig. 3} Real and imaginary parts of the conductivity for $\alpha = 1, 0.6$ and $0.1$  superconducting state of {\em single-layer} superconductor calculated as described in refs\cite{Ioffe99,Ioffe00}.}
\vspace{0.25cm}

Fig 3b shows Re and Im $\sigma$ for a typical `intermediate' fluctuation parameter $\alpha = 0.6$. The crucial effects of reducing $\alpha$ from unity are seen to be (1) elimination of the type-II coherence factor effect at the gap edge (so within this model the Re$\sigma$ acquires a step at the gap edge) and (2) a corresponding negative divergence of Im$\sigma$ at $\omega = 2 \Delta$, leading to two zero-crossings near $2\Delta$. The $\omega > 2 \Delta$ zero crossing combined  with the smallness of Re $\sigma$ gives rise to a sharp peak in $\tau^{-1}$ (appearing to be at the gap edge on the scale of the plot in Fig 1) and the $\omega < 2 \Delta$ zero crossing leads in this model to a weak delta-function contribution to $\tau^{-1}$ as  Re$\sigma$ is zero in that region. At a  higher frequency the Im part crosses over to the  negative  side but  $\tau^{-1}$ continues to decrease monotonically. 
Fig 3c shows Re and Im $\sigma$ for strong fluctuations, $\alpha = 0.1$. The increase in fluctuations substantially weakens the $1/\omega$ divergence in Im$\sigma$ and strengthens the negative divergence at the gap edge, eliminating the $\omega > 2 \Delta$ zero crossing and moving the $\omega < 2 \Delta$ zero crossing to lower frequencies, substantially strengthening the delta function at low frequencies. In the {\em pseudogap} regime ($\alpha = 0$, not shown) the 
$\omega < 2 \Delta$ zero crossing (and the associate delta function in $\tau^{-1}$ ) moves to $\omega = 0$.


We will now  outline  the new features that are introduced in the analysis of Re$\sigma ^{-1}$ when we consider the generalization of \cite{Ioffe99,Ioffe00} to the {\em bi-layer} case. A detailed discussion of the c-axis conductivity for the {\em bi-layer} case is given elsewhere \cite{SMI}. The essential point is that in a {\em bi-layer} system a uniform applied field may lead to a  non-uniform charge distribution within a unit cell which in turn produces electrochemical potential differences which affect charge flows. These ``local-field effects'' give  rise for example to the bilayer plasmon feature observed in many {\em bi-layer} materials such as $YBa_{2}Cu_{3}O_{6+x}$ \cite{Marel98}. 

We consider here a {\em bi-layer} structure with a unit cell consisting of two identical superconducting planes separated by a distance $d_1$, and with a plane in one cell separated from the nearest plane in another unit cell by distance $d_2$. The approach used in \cite{Ioffe99,Ioffe00} yields 
\begin{equation}
\sigma_{bilayer}(\omega )=\frac{\sigma _{1}\sigma _{2}-i\omega (\sigma
_{1}\widetilde{d}%
_{1} +\sigma _{2}\widetilde{d}_{2}  )/C}{\sigma _{1}\widetilde{d}_{2}+\sigma _{2}\widetilde{d}%
_{1}-i\omega /C}  
\label{sigmabilayer}
\end{equation}
with $\widetilde{d}_{1,2}=d_{1,2}/(d_{1} + d_{2})$ and $\sigma_{1, 2}$ the conductivities appropriate to a pair of planes separated by distance $d_{1, 2}$ and C a blockade parameter expressing the electrochemical potential difference created by a charge imbalance within the unit cell. We assume the two conductivities to be the same up to a scale factor $b = \sigma _{2}/\sigma _{1}$ coming from differences in the hopping amplitudes. Eq \ref{sigmabilayer} implies a characteristic frequency scale  
\begin{equation}
\omega^{*}= C \sigma _{1}(\widetilde{d}_{2}+ b \widetilde{d}_{1}).
\label{omegabilayernorm}
\end{equation}
For reasons discussed earlier, in our discussion we will assume that the normal state conductivities $\sigma_{1,2}(\Delta = 0)$ are frequency independent and use these  to define $\omega^{*}$.

Fig 4 plots the Re and Im $\sigma_{bilayer}$ for the normal($\Delta = 0$) state and for $\alpha = 0.1$ superconducting  state with $\omega^{*} < 2\Delta$. The low-frequency suppression in the normal state caused by the local field effects gives rise to a low-frequency upturn in $\tau^{-1}$. 
The superconductivity-induced changes which have been discussed in detail for the {\em single-layer} case also apply to the {\em bi-layer} case. But new features may occur in the {\em bi-layer} case. In particular, for some parameter ranges including those used in Fig 4b, the local field effects lead to a bilayer plasmon (divergence in $\sigma_{bilayer}$) at $\omega < 2\Delta$. This produces an additional zero crossings in Im$\sigma$, shown as the lower-frequency circle in Fig 4b. 
The second zero-crossing at a higher frequency (again circled) has it's origin in the divergence at the gap edge discussed earlier for the {\em single-layer} case (compare Figs 3c and 4b). 

\vspace{0.05cm} 
\centerline{\epsfxsize=3truein \epsfbox{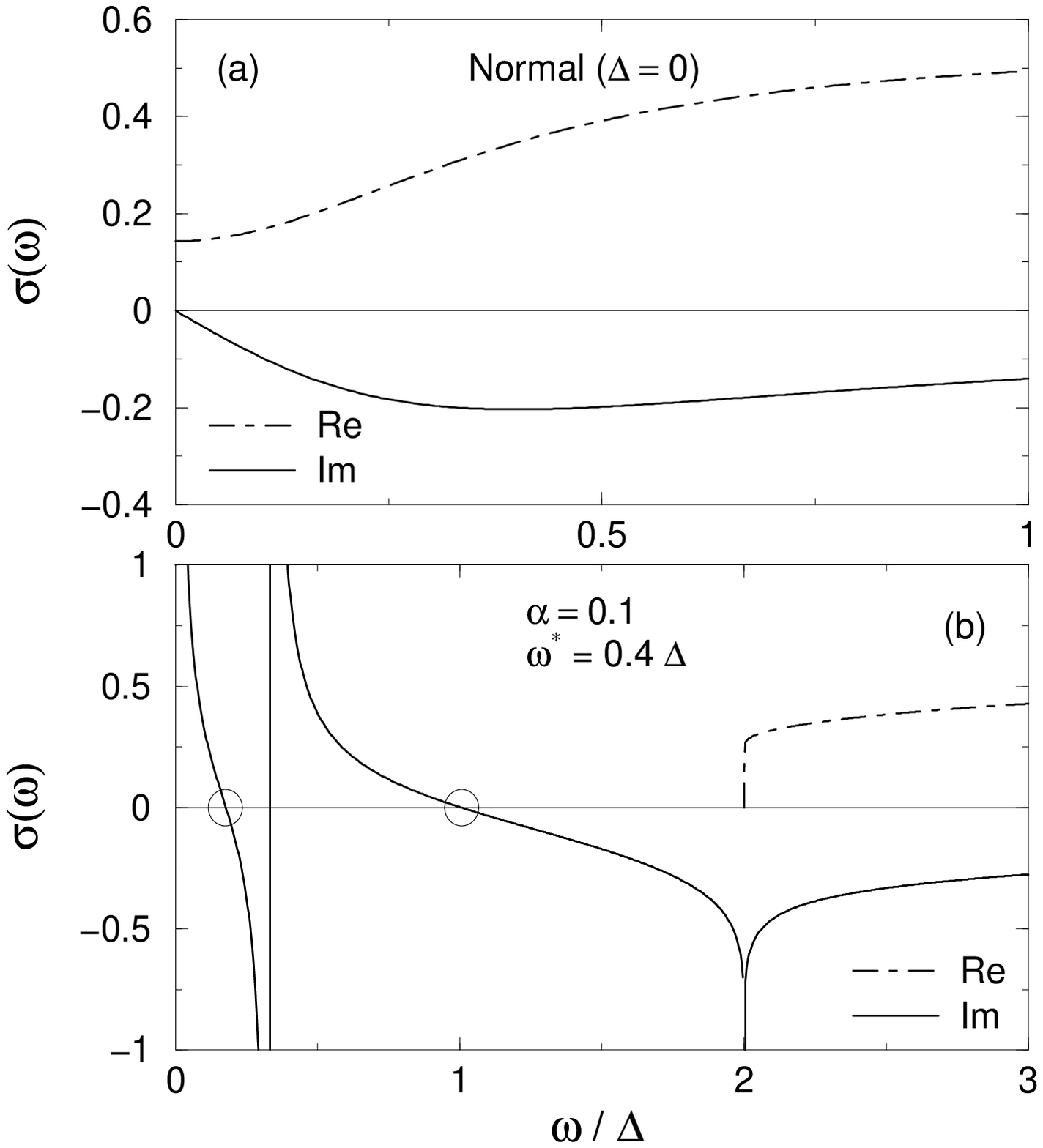}}
{\footnotesize \textbf{Fig. 4} Real and imaginary parts of the conductivity of a {\em bi-layer} superconductor with $b = 0.075$ calculated as described in ref\cite{SMI} for the normal state ($\Delta = 0$) and for an $\alpha = 0.1$  superconducting state with $\omega^{*} = 0.4 \Delta$.}
\vspace{0.25cm}

 Corresponding to the two zero-crossings (Fig 4b) there are two delta-functions in Re$\sigma ^{-1}$.  These two delta-functions are obtained also for other values of $\alpha$ but their relative weights vary with the value of $\alpha$. The form of $\tau^{-1}$ normalized to $\omega \rightarrow \infty$ limit for the case when the bilayer plasmon lies inside the gap for all non-zero values of $\alpha$, coincides exactly with that for the corresponding {\em single-layer} case as shown in Fig 1. 

\vspace{0.25cm} 
\centerline{\epsfxsize=3truein \epsfbox{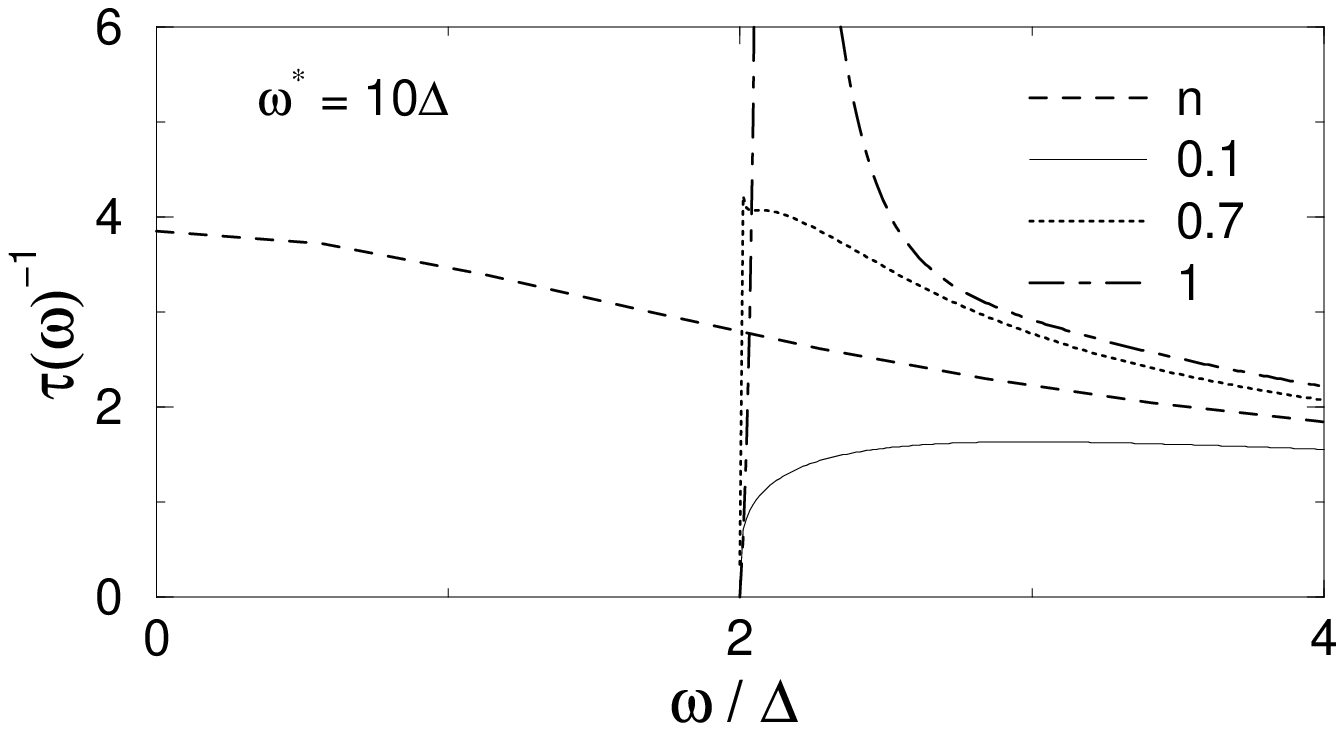}}
{\footnotesize \textbf{Fig. 5} Optical scattering rate $\tau^{-1}$ (Eq \ref{exp}) for a {\em bi-layer} superconductor with $b = 0.075$ and an $\omega^{*} = 10 \Delta$ normalized to $\omega \rightarrow \infty$ limit, in normal state (dashed line) and for superconducting state for different values of fluctuation parameters $\alpha = 0.1, 0.7$ and 1. For superconducting state only the contribution for $\omega >2\Delta $ is shown.}
\vspace{0.25cm}

The detailed structure of zero-crossing and the form of conductivity varies with the ratio of the gap to the cross-over frequency  $\omega^{*}$ (Eq \ref{omegabilayernorm}), thus changing the form of $\tau^{-1}$. For $\omega^{*} >> 2\Delta$, and $\alpha =1$ case for example, the {\em bi-layer} structure introduces a  zero-crossing of Im$\sigma$ just above the gap, which combined with the smallness of the Re$\sigma$ gives rise to a strong divergence in $\tau^{-1}$ as shown in Fig 5.

The sum-rule arguments can be easily extended to the {\em bi-layer} case and we get the sum-rule as given by \ref{finalsum} for the {\em single-layer} case but with $f(\alpha)$ replaced by 
\bea
f_{bilayer}(\alpha) = \frac{1}{\widetilde{d}_{1}+ b \widetilde{d}_{2}}\frac{\Delta \pi \left( 1-\alpha \right) }{2\sigma _{1}(\Delta = 0)}.
\eea


In conclusion, we have studied the behavior of ``optical scattering rate'' Re$\sigma^{-1}$ in a model for the c-axis conductivity of high temperature superconductors. The presence of quantal and thermal fluctuations in the phase of the superconducting order parameter is shown to lead to a non-intuitive increase in the integrated area of the ``scattering rate'', Re$\sigma^{-1}(\omega)$. This increase is a net result of the contributions from the $\omega < 2 \Delta$ and  $\omega > 2 \Delta$ regions   which depend sensitively on the strength of the phase fluctuations. In the strong phase fluctuation limit, Re$\sigma^{-1}(\omega > 2 \Delta)$  is decreased below its $\Delta = 0$ value whereas for weak phase fluctuations it is increased. While the variations in Re$\sigma^{-1}$ are seen to be a useful probe of order parameter phase fluctuations, the large contributions at $\omega < 2 \Delta$ coming from zero-crossings of Im$\sigma$ suggest that caution is required in interpreting Re$\sigma^{-1}$ as a scattering rate.

We thank D. Basov for helpful discussions and van D. van der Marel for correcting a mistake in the paper. We thank NSF DMR 00081075 for support.

\end{multicols}
\end{document}